\documentclass[aps,prl,reprint,a4paper,superscriptaddress,numbers,longbibliography,showpacs,showkeys]{revtex4-1}

\usepackage[pdftex]{hyperref,color,graphicx}
\usepackage{amsfonts,amssymb,amsmath}
\usepackage[separate-uncertainty=false]{siunitx}
\usepackage[english]{babel}
\usepackage[utf8]{inputenc}
\usepackage[T1]{fontenc}

\usepackage{xspace}

\hypersetup{
    pdftitle = {Low-entropy states of neutral atoms in polarization-synthesized optical lattices},
    pdfauthor = {Carsten Robens, Jonathan Zopes, Wolfgang Alt, Stefan Brakhane, Dieter Meschede, and Andrea Alberti},
	colorlinks=true,linkcolor=blue,citecolor=blue,filecolor=blue,urlcolor=blue
  }

\newif\ifusebibfile
\usebibfiletrue

\newcommand{\figref}[2]{\hyperref[#1]{\ref{#1}(#2)}}

\def\ket#1{\mathinner{|{#1}\rangle}}

\newcommand{\spinup}{\ket{{\uparrow}}}
\newcommand{\spindown}{\ket{\downarrow}}

\def\PSOLAS{PSOLAS\xspace}

\DeclareSIUnit\gauss{G}
\DeclareSIUnit\centimeter{cm}

\addto\captionsenglish{}
\addto\captionsenglish{}

\renewcommand\textemdash{\leavevmode\unskip\kern0.8pt\rule[0.19\baselineskip]{8pt}{0.4pt}\kern1pt\ignorespaces}

\renewcommand{\section}[1]{\textit{#1}.\textemdash}

\begin{document}
\textheight=24.1cm
\selectlanguage{english}

\title{Low-entropy states of neutral atoms in polarization-synthesized optical lattices}

\author{Carsten Robens$^{1,\dagger}$, Jonathan Zopes$^{1,\dagger}$, Wolfgang Alt$^{1}$, Stefan Brakhane$^{1}$, Dieter Meschede$^{1}$, and Andrea Alberti} \email{alberti@iap.uni-bonn.de} \thanks{$^{\dagger}$Both authors contributed equally to this work.} \affiliation{Institut für Angewandte Physik, Universität Bonn, Wegelerstr.~8, D-53115 Bonn, Germany}

	\begin{abstract}
	We create low-entropy states of neutral atoms by utilizing a conceptually new optical-lattice technique that relies on a high-precision, high-bandwidth synthesis of light polarization.
	Polarization-synthesized optical lattices provide two fully controllable optical lattice potentials, each of them confining only atoms in either one of the two long-lived hyperfine states.
	By employing one lattice as the storage register and the other one as the shift register, we provide a proof of concept {using four atoms} that selected regions of the periodic potential can be filled with one particle per site.
	{We expect that our results can be scaled up to thousands of atoms by employing a atom-sorting algorithm with logarithmic complexity, which is enabled by polarization-synthesized optical lattices.}
	Vibrational entropy is subsequently removed by sideband cooling methods.
	Our results pave the way for a bottom-up approach to creating ultralow-entropy states of a many-body system.

\end{abstract}
\maketitle

\section{Introduction}
Compared to other quantum systems, optical lattice potentials stand out for being naturally scalable.
They offer thousands of sites, arranged in periodic arrays, in which quantum particles such as atoms can be confined and manipulated \cite{Bloch:2012}.
The idea of employing the myriad of sites available as a well-controlled Hilbert space has influenced modern research frontiers ranging from quantum metrology \cite{Katori:2011}, quantum information processing \cite{Jaksch99,Brennen:1999,Raussendorf:2001,Lee:2013,Xia:2015,Wang:2016}, discrete-time quantum walks \cite{Karski09}, up to quantum simulations of strongly correlated condensed-matter systems \cite{Jaksch:1998,Greiner:2002,Lewenstein:2012} with single lattice-site resolution \cite{Bakr:2010,Sherson10}.
Substantial experimental effort has recently been devoted to creating low-entropy states of atoms in the lattice, with each site being occupied by an integer number of atoms.
Low-entropy states play an essential role in a host of quantum applications including the creation of highly entangled cluster states for quantum information processing \cite{Mandel04}, investigation of Hong-Ou-Mandel-like quantum correlations in many-body systems
\cite{Kaufman14,Islam:2015},
and the quantum simulation of quantum spin liquids in frustrated systems \cite{Leggett:2006,Esslinger:2010}.

To date, the approach that has proven most effective to generate low-entropy states in optical lattices relies on a Mott insulator phase \cite{Jaksch:1998,Greiner:2002}.
This is denoted as a top-down approach since ultracold atoms, due to interactions, self-organize in domains with integer filling factors.
Other approaches \cite{Fung:2016} relying only on laser cooling techniques have recently demonstrated filling factors beyond the one-half limit imposed by inelastic light-assisted collisions \cite{DePue:99,Fuhrmanek:2012}, though without providing a fully deterministic method.
In contrast, a bottom-up approach generating arbitrary low-entropy states from individual atoms has long been desired \cite{Jaksch99,Brennen:1999}, yet never been experimentally realized.
In this Letter, we demonstrate a bottom-up approach to generate arbitrary atom patterns, including unity filling of lattice sites, in a one-dimensional (1D) optical lattice.
Inspired by the seminal work by Jaksch \emph{et al.}~\cite{Jaksch99} proposing spin-dependent optical lattices to control individual atoms' positions, our work realizes the atom-sorting scheme proposed by Weiss \emph{et al.}~\cite{Weiss04}.
The experimental challenge consists in developing spin-dependent optical lattices able to shift atoms by any amount of lattice sites conditioned to their spin state.
Previous implementations \cite{Mandel03,Steffen12} of spin-dependent optical lattices were limited  to only relative displacements of the two spin components and to relative shift distances of one site at most.
To overcome these limitations, we have devised a scheme for spin-dependent transport based on a high precision, large bandwidth synthesizer of polarization states of light.
Hence, we refer to our new implementation of spin-dependent optical potentials as polarization-synthesized (PS) optical lattices.
{PS optical lattices allow us to reposition individual atoms with a precision of $\SI{1}{\angstrom}$, reducing thereby the positional entropy of a randomly distributed ensemble to virtually zero.}
{This is in stark contrast to the atom-sorting technique formerly demonstrated by our group \cite{Miroshnychenko06}, whose positioning precision was limited to about five sites.}
 In addition, the novel approach
requires no post-selection, which have limited the success rates in earlier efforts to create ordered patterns from a thermal ensemble \cite{Schrader04,Karski10-ap}.

\begin{figure}[t]
	\includegraphics*[width=8.8cm]{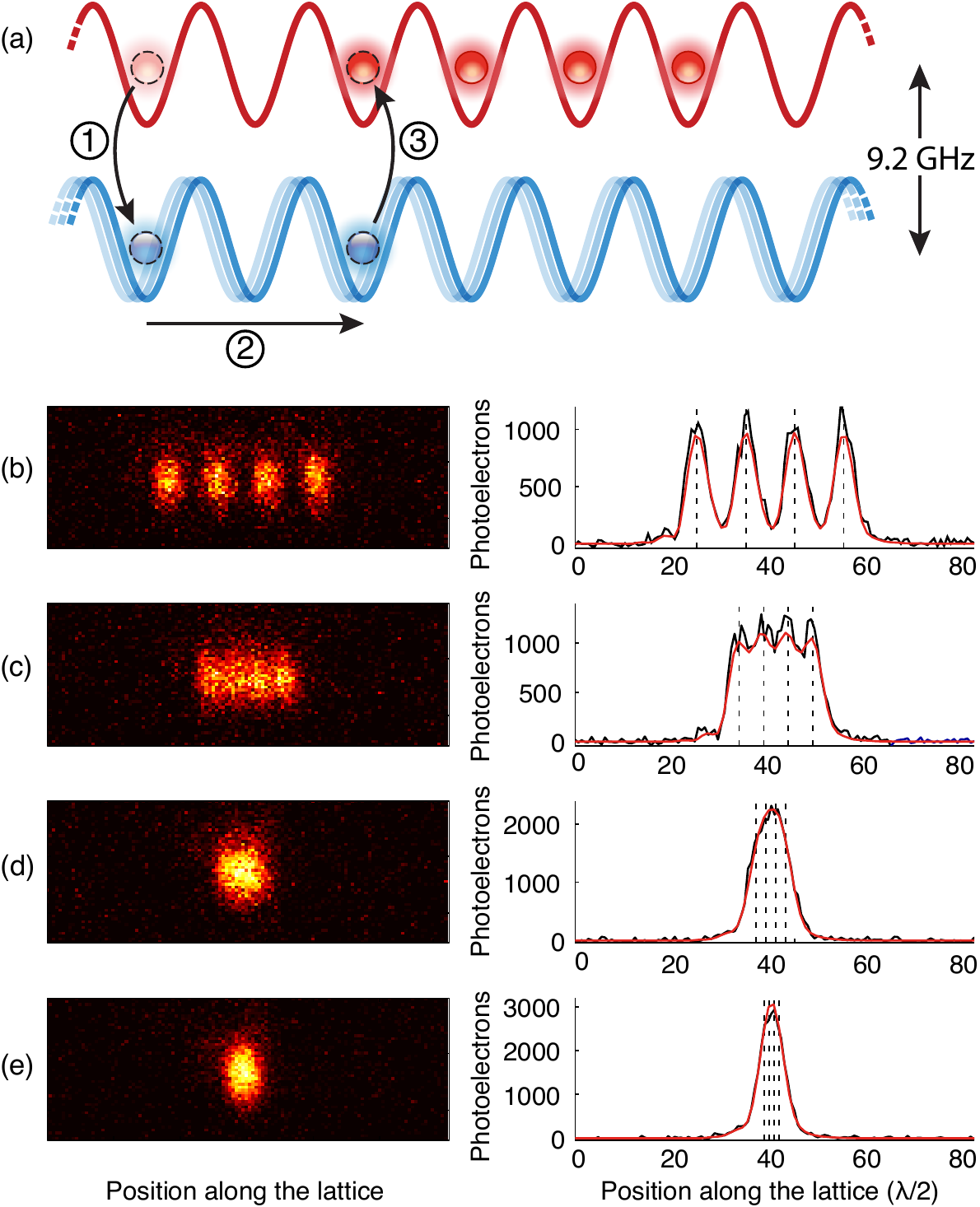}
	\caption{\label{fig:unity-filling}
	Atom sorting in polarization-synthesized optical lattices. (a) Central building block of the atom-sorting procedure: (1) the leftmost atom (marked by dashed circle) is transferred from the \emph{storage register} (upper lattice) into the \emph{shift register} (lower lattice) by a microwave pulse, (2) transported by two sites to the right by shifting the lower lattice, (3) and transferred back into the storage register.
	(b--e) From top to bottom, four atoms deterministically placed at equidistant separations of $ d_\text{target}= ( 10, 5, 2, 1 )$ lattice sites.
	Left panels: recorded single-shot fluorescence images.
	Right panels: vertically integrated distributions {(black lines)} with the fitted intensity profiles (red curves) and the reconstructed positions (vertical dashed lines).
	}
\end{figure}

\section{Atom sorting}
The principal result of this work is shown in Fig.~\ref{fig:unity-filling}:
four ${}^{133}$Cs atoms from a dilute thermal ensemble are rearranged into a predefined, ordered distribution inside a 1D optical lattice.
The atom-sorting procedure works akin to Maxwell's demon.
In essence, an automated feedback-based experimental setup acquires the initial location of atoms through fluorescence imaging with single site precision, and it uses this information to subsequently shift the atoms, one by one, to form the desired pattern.
As illustrated in Fig.~\figref{fig:unity-filling}{a}, two spatially overlapped optical lattices,
  \begin{align}
	\label{eq:spinup}
	U_\uparrow(x,t) &= U^{0}_{\uparrow} \cos^2\hspace{-1pt}\{k_\text{L}[x-x_\uparrow(t)]\}\,,\\
	\label{eq:spindown}
		U_\downarrow(x,t) &= U^{0}_{\downarrow}\cos^2\hspace{-1pt}\{k_\text{L}[x-x_\downarrow(t)]\}\,,
\end{align}
with identical lattice constant ($\pi/k_\mathrm{L}$) are used to sort atoms.
The first lattice is kept fixed, serving as a \emph{storage register} for atoms in the hyperfine state $\spinup = \ket{F=4,m_F=4}$, while the other one is mobile, providing a \emph{shift register} for atoms in $\spindown = \ket{F=3,m_F=3}$.
A digitally programmable polarization synthesizer, as will be detailed later, gives us full independent control of both lattice depths \raisebox{0pt}[0pt][0pt]{$U_{s}$} ($s \in \{\uparrow,\downarrow\}$) as well as of the lattice positions $x_s(t)$, which are varied in time to shift the atoms.
We choose deep lattices, of the order of a thousand recoil energies, to allow fast transport on the timescale of ten microseconds, while preventing intersite tunneling.
For each atom we intend to reposition, we flip its spin, $\ket{{\uparrow}}\rightarrow \ket{{\downarrow}}$, using a position-resolved microwave pulse \cite{Karski10-ap,Johanning:2009}, which transfers it into the shift register.
Once the atom is repositioned by translating the shift register, it is transferred back into the storage register through optical pumping.
The fluorescence images in Fig.~\figref{fig:unity-filling}{b\hspace{1pt}\rule[0.19\baselineskip]{4pt}{0.4pt}\hspace{1pt}e} show the final distribution of atoms for four different target patterns, including unity filling of a region of the lattice.
Between fluorescence images, several sorting operations are carried out with no need to continuously monitor the positions of atoms.
If errors are detected in the final distribution (e.g., imperfect spin-flips, wrong position reconstruction, atom losses), a feedback control system attempts to correct them.

\section{Experimental apparatus}
A small ensemble of cesium atoms is captured from the background vapor into a magneto-optical trap, and subsequently transferred into an 1D optical lattice produced by linearly polarized light at the wavelength $\lambda_\mathrm{L}=2\pi/k_L=\SI{866}{\nano\meter}$.
{Lifetime of atoms due to collisions with background gas is about \SI{360}{\second}.}
The lattice depth is chosen equal for both spin species, $U^0_\uparrow=U^0_\downarrow\approx\SI{75}{\micro\kelvin}$, and significantly larger than the atoms' temperature, which is about $\SI{8}{\micro\kelvin}$ after molasses cooling.
The loading procedure is adjusted to spread the atoms along the lattice with an average separation of around 20 lattice sites.
Optical pumping initializes atoms in $\spinup$ state with $> \SI{99}{\percent}$ efficiency using a $\sigma^+$-polarized laser.
To detect the atoms' positions, we acquire fluorescence images with $\SI{1}{\second}$ illumination time using an electron-multiplying CCD camera.
We employ a superresolution-microscopy technique \cite{Alberti2015} to resolve in real time the individual atoms beyond the diffraction limit of around four sites, as can be seen comparing Fig.~\figref{fig:unity-filling}{c} and \figref{fig:unity-filling}{d}.
The local addressing of individual atoms is achieved through spectrally narrow Gaussian-shaped microwave pulses ($\SI{7}{\kilo\hertz}$ RMS width) in combination with a weak magnetic field gradient ($\SI{11.6}{\gauss/\centimeter}$) along the direction of the optical lattice \cite{Karski10-ap}.
For the sorting procedure, we select atoms isolated by more than \num{20} sites to ensure a probability $<\SI{1}{\percent}$ that local addressing pulses spin-flip a neighboring atom.
We choose adiabatic, sinusoidal ramps to transport the addressed atoms in the shift register in approximately \SI{1}{\milli\second}, much shorter than the longitudinal spin relaxation time of \SI{100}{\milli\second} due to inelastic scattering of the lattice photons.
We pump atoms back into the storage register by 2-\si{\milli\second} optical pumping.
Atoms in excess are removed from the lattice by first spin-flipping the sorted atoms into the shift register, and by subsequently applying a resonant pulse with the $F=4 \rightarrow F'=5$ transition, pushing atoms in the $\spinup$ state out of the optical lattice, while not affecting atoms in the $\spindown$ state.
Our deep optical lattice, combined with a superimposed blue-detuned doughnut-shaped trap, yields longitudinal and transverse trapping frequencies of about $\omega_\parallel\approx2\pi\times \SI{110}{\kilo\hertz}$ and  $\omega_\perp\approx2\pi\times \SI{20}{\kilo\hertz}$, well above the recoil frequency $2\pi\times \SI{2}{\kilo\hertz}$ of cesium.
The large trapping frequencies allow us to employ, subsequent to the atom-sorting procedure, microwave \cite{Forster:2009} and Raman \cite{Kaufman12} sideband cooling to cool atoms in the longitudinal and radial direction, respectively, achieving a ground state occupation of $\SI{99}{\percent}$ along the axial direction and of $\SI{80}{\percent}$ along each of the two transverse directions.

\begin{figure*}[t]
	\includegraphics*[width=17.7cm]{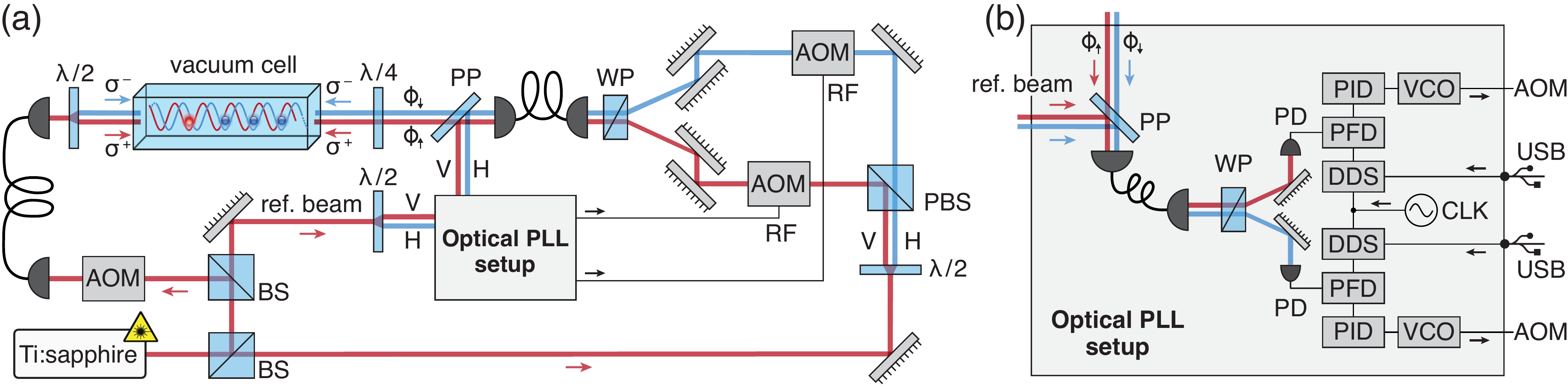}
	\caption[]{Schematic illustration of the experimental setup for polarization-synthesized optical lattices.
	(a) The linearly polarized output of a Ti:sapphire laser is split by beamsplitters (BS) into the reference beam, which is used for the optical phase-locked loops (PLLs), and the beams forming the lattice in the vacuum cell.
	While the polarization of the left lattice beam is static and linear, the polarization of the right lattice beam is synthesized by overlapping two beams of opposite circular polarizations.
	The latter are combined by a Wollaston prism (WP) in linear polarization basis (vertical V, horizontal H), spatially mode matched by a polarization maintaining optical fiber (high polarization extinction ratio $>\SI{50}{\decibel}$ \cite{Sears:1990}), and transformed into circular polarizations by a $\lambda/4$ plate.
	A fraction of light is diverted by a pick-up plate (PP) into the optical PLL setup, which  controls the optical phases $\phi_\uparrow$ and $\phi_\downarrow$ by feeding RF signals back to the acousto-optic modulators (AOMs).
	(b) Optical PLL setup:
		the diverted light is overlapped with a common reference beam.
	The resulting beat signals are independently recorded by fast photodiodes (PD) after the WP.
	The phase of each beat signal is compared with a RF reference signal (DDS)
		using a digital phase-frequency discriminator (PFD),
		and fed to a PID controller (\SI{10}{\mega\hertz} bandwidth), which steers the corresponding AOM through a voltage-controlled oscillator (VCO).
	The DDS RF sources are phase referenced to the same \SI{400}{\mega\hertz} clock signal (CLK) and interfaced via USB with a computer.
	Three additional control-loop setups (not shown) independently regulate the intensity of each lattice beam by controlling the RF power of the corresponding AOM. \label{fig:transport-setup}
	}
																		\end{figure*}

\section{Polarization-synthesized optical lattices}
The key element in realizing the spin-dependent optical-lattice potentials shown in  Eqs.~(\ref{eq:spinup}) and (\ref{eq:spindown}) are two superimposed, yet independently controllable optical standing waves with opposite circular polarization, $\sigma^+$ and $\sigma^-$.
For both standing waves we choose a so-called magic wavelength $\lambda_L$ of cesium, allowing
atoms in $\spinup$ and $\spindown$ state to be trapped in the maximum-intensity regions of the $\sigma^+$- and $\sigma^-$-polarized light field, respectively \cite{Belmechri:2013}.
Such a wavelength exists because of the different AC vector polarizability of the two internal states \cite{Deutsch:1998}, and was already employed in earlier implementations of spin-dependent optical lattices (e.g., Refs.~\cite{Mandel03,Steffen12}).
However, these implementations permitted only relative displacements and, most importantly, maximum shift distances of one lattice site, thereby precluding the possibility of sorting randomly distributed atoms into predefined patterns.
In contrast, PS optical lattices entirely overcome these limitations by relying on two fully independent optical standing waves.
In order to create the standing waves, we let two co-propagating laser beams with opposite circular polarization each interfere with a linearly-polarized, counter-propagating beam, as illustrated in Fig.~\figref{fig:transport-setup}{a}.
We employ an optical fiber to ensure that the resulting standing waves are perfectly matched to the same transverse mode, and thereby that atoms in both spin states, $\spinup$ and $\spindown$, experience an identical transverse potential.
Transverse-mode filtering is essential to ensure long spin-coherence times for spectrally-narrow coherent pulses (e.g., spin flips for single-atom addressing), or else thermal atoms would undergo inhomogeneous spin dephasing in a few microseconds due to a strong differential light shift \cite{Alberti14}.

While in the transverse directions the two standing waves are perfectly overlapped, they are free to slide with respect to each other in the lattice direction.
The position of each standing wave must be controlled with interferometric precision to ensure that atoms are shifted with single-site precision, and that no motional excitation is created when atoms are transferred between the storage and shift registers \cite{Belmechri:2013}.
We achieve this by employing two independent optical phase-locked loops (PLLs) that actively stabilize the phases of each circularly-polarized beam, $\phi_{\uparrow}$ and $\phi_{\downarrow}$, with respect to a common optical reference beam.
As shown in Fig.~\figref{fig:transport-setup}{b}, each optical phase $\phi_s$ is referenced to a low-phase-noise RF reference signal (DDS).
Varying the phase of the RF signals according to a digitally programmed profile allows us to independently steer $\phi_{s}$, and thereby the position of the respective optical potential $U_s$:
\begin{equation}\label{eqn:PhasePos}
x_{s}(t) = \frac{\lambda_{\mathrm{L}}}{2} \frac{ \phi_{s}(t)}{ 2\pi}\,.
\end{equation}
{A measurement of the relative phase noise $\Delta \phi=\phi_\uparrow-\phi_\downarrow$ yields an uncertainty of $\SI{0.1}{\degree}$~\cite{Robens:2016pol}, which translates into a jittering of the relative position of $\Delta x=\SI{1.20}{\angstrom}$.
This is more than two orders of magnitude smaller than the extent of the atomic wave function in the vibrational ground state (20\,nm) \cite{RelativePrecisionNote}.}
Moreover, to attest the reliability of the spin-dependent transport operations, we shifted spin-polarized atoms
using a single transport operation, chosen about $\SI{1}{\milli\second}$ long, over a distance varying from a few to one hundred sites.
We measure a success rate of $\SI{97.4(3)}{\percent}$, nearly independent of the distance.
We attribute the remaining unsuccessful events to optical pumping errors ($\SI{0.4}{\percent}$), spin-flips during transport ($\SI{0.6}{\percent}$), and position reconstruction errors ($\SI{1.6}{\percent}$).
This dramatically differs from previously reported transport efficiencies, decreasing exponentially with the transport distance.
Even with the best reported efficiency of $\SI{99}{\percent}$ per shift operation \cite{Karski:2011}, transport efficiency over $20$ sites had never exceeded $0.99^{2\cdot 20}\approx \SI{67}{\percent}$.

{\section{Thousands of atoms}
While low-entropy states comprising as few as four atoms already suffice to study a wide range of few-body phenomena \cite{Kaufman14,Islam:2015,Murmann:2015,Murmann:2015b}, it is important to discuss how the atom-sorting scheme based on PS optical lattices can be extended to much larger numbers of atoms, thus providing a bottom-up pathway to many-body physics.
Very recently, different schemes based on movable optical tweezers \cite{Barredo:2016,Endres:2016} allowed sorting about 50 atoms into predefined positions by rearranging them one by one;
the optical-tweezers atom-sorting scheme appears particularly suited for Rydberg physics where atoms sit at a relatively large distance from each other.
Compared to these recent results, fewer atoms are sorted in Fig.~\ref{fig:unity-filling} because of the limited addressing resolution (20 sites) of the present apparatus (see Supplemental Material).
However, we expect that with a higher addressing resolution \cite{Robens:2016mbg} PS optical lattices allow sorting even thousands of atoms into arbitrary target patterns.

To that purpose, we propose a new atom-sorting algorithm based on PS optical lattices, which rearranges $N$ atoms using a number of operations of the order of $\log N$.
The logarithmic complexity is enabled by two properties unique to PS optical lattices, namely their ability to shift atoms (1) spin dependently and (2) by any arbitrary number of sites;
moreover, its logarithmic complexity also holds for two-dimensional (2D) PS optical lattices, which have been recently proposed in Ref.~\citenum{Groh:2016}.
Hitherto, the best algorithm \cite{Vala05} for sorting atoms in a 2D array requires a larger number of operations of the order of $N^{1/2}$, because only one-lattice-site shifts are used instead of property (2).

In essence, the PS-optical-lattice atom-sorting (\PSOLAS) algorithm proposed here iterates four steps:
Step 1 identifies among the atoms not yet sorted patterns of atoms that best match the distribution of defects, i.e., the empty sites to be filled with one atom; this step requires acquiring the positions of the atoms through a fluorescence image \cite{FluorescenceImages}.
Step 2 transfers the identified atoms from the storage register into the shift register; this step can be performed in parallel by optically addressing atoms using a spatial light modulator \cite{Gauthier:2016,Nogrette:2014} or serially using a beam deflector \cite{Weitenberg11,Barredo:2016}.
Step 3 shifts the whole pattern of selected atoms, in parallel, to fill the defects; this steps is the crucial one, which is enabled by the properties (1) and (2) of PS optical lattices.
Step 4 transfers the shifted atoms into the storage register by optical pumping.
An illustrative demonstration of \PSOLAS algorithm to fill a square region of a 2D optical lattice is provided in the Supplemental Material.

\begin{figure}[t]
	\includegraphics*[width=\columnwidth]{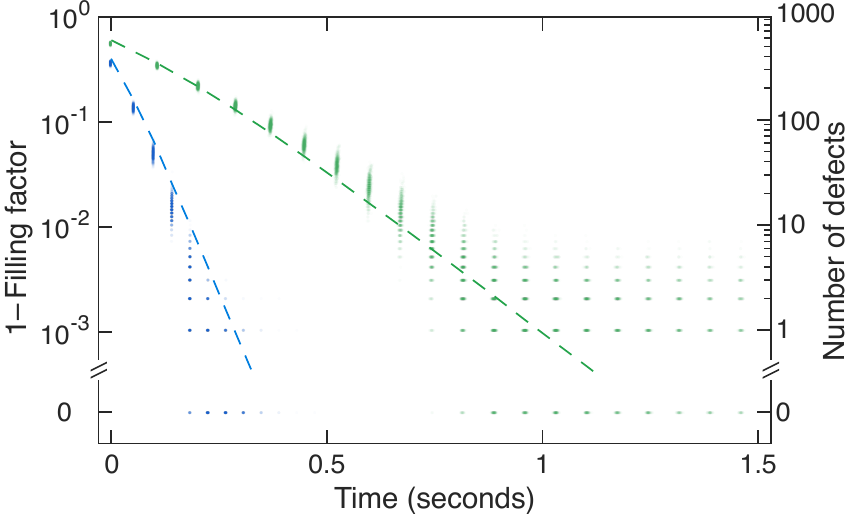}
	\caption{\label{fig:Montecarlo}
	{Monte Carlo simulation of \PSOLAS algorithm filling a square region of a 2D optical lattice with 961 atoms.
	Upper and lower datasets refer to scenarios (A) and (B) discussed in the text.
	Dashed lines denote, for each scenario, the exponential law derived in the text.
	 	The data show that with $\SI{95}{\percent}$ probability, defect-free unity filling is reached in less than $\SI{0.3}{\second}$ (A) and $\SI{2.2}{\second}$ (B).
	In scenario (A), deviations from the exponential law are attributed to the conservative conditions.
	Nevertheless, the scaling law remains exponential, with a reduced effective filling probability.
	}
	}
\end{figure}

Since the duration of each step is independent from the number of sorted atoms for a broad parameter range, the overall time required by \PSOLAS is determined by the number of iterations.
The latter is simply estimated by considering that each iteration fills, on average, a fraction $\alpha$ of the defects in the target pattern, where $\alpha$ denotes the initial filling probability of a lattice site.
In reality, because the algorithm searches for the best matching pattern of atoms to shift, the fraction of filled defects per iteration is generally higher than $\alpha$.
Hence, the number of defects after $n$ iterations amounts to less than $(1-\alpha)^{1+n}$, meaning that
to attain a number of defect of the order of $\mathcal{O}(1)$, about $\mathcal{O}(\log N)$ iterations are required.
To validate this scaling law under realistic conditions, we carried out Monte Carlo simulations in two scenarios representing (A) conservative and (B) state-of-the-art conditions; the conditions of both scenarios are derived from  individual results demonstrated either in our or other laboratories.
In both scenarios, \PSOLAS aims to fill a square target pattern of $31\times 31$ sites by ``tapping'' into the atoms stored in a larger region of $100\times 100$ sites.
Scenario (A) and (B) rely on $\SI{80}{\percent}$ and $\SI{95}{\percent}$ \cite{Weitenberg11} single-site addressing efficiency,
and the filling probability $\alpha$ is chosen equal to $\SI{40}{\percent}$ and $\SI{60}{\percent}$ \cite{Fung:2016}, respectively (all  parameters are summarized in the Supplemental Material).
As shown in Fig.~\ref{fig:Montecarlo}, we find that even in the conservative scenario (A), about 1000 atoms can be sorted in a time of about $\SI{1}{\second}$.
}

\section{Conclusions}
In this paper, we demonstrated a bottom-up approach to the generation of low-entropy states of ultracold atoms in optical lattices.
Our work demonstrates that arbitrary filling patterns with virtually zero entropy can be realized experimentally.
The key to our sorting procedure is the development of PS optical lattices, which provide us with a new set of operations for the control of atoms depending on their spin orientation.
Presently, the entropy of our prepared states is limited by the vibrational entropy \cite{Olshanii:2002} due to the limited efficiency ($\SI{60}{\percent}$) of the sideband cooling into the three-dimensional vibrational ground state.
A tighter optical confinement of atoms shall enable significantly higher efficiencies.
{The construction of a 2D PS optical lattice is underway \cite{Groh:2016} in a new experimental apparatus, which additionally features a high optical resolution objective lens for optical single-site addressing \cite{Robens:2016mbg}; this should allow us to demonstrate \PSOLAS with thousands of atoms.}

\begin{acknowledgments}We thank A.\ Hambitzer for contributing to the experimental apparatus and S.\ Hild, A.\ Steffen, G.\ Ramola, M.\ Tarallo, J.\ M.\ Raimond and C.\ Kollath for insightful discussions.
We acknowledge financial support from the NRW-Nachwuchsforschergruppe ``Quantenkontrolle auf der Nanoskala'', the ERC grant DQSIM, the EU SIQS project, and the Deutsche Forschungsgemeinschaft SFB project OSCAR.
C.R.\ acknowledges support from the Studienstiftung des deutschen Volkes, and C.R., S.B., and J.Z. from the Bonn-Cologne Graduate School.
\end{acknowledgments}

\bibliographystyle{apsrev4-1}
\bibliography{references}

\begin{thebibliography}{51}
\bibitem [{Rel()}]{RelativePrecisionNote}%
  \BibitemOpen
  \bibinfo {note} {The relative position jitter must be much smaller then the
  harmonic oscillator length $\sqrt{\hbar/(2 m\omega_\parallel)}$, with $m$
  being the mass of the atom.}\BibitemShut {Stop}%
\bibitem [{Flu()}]{FluorescenceImages}%
  \BibitemOpen
  \bibinfo {note} {Note that a fluorescence image is not strictly required at
  each iteration since atoms can also be sorted blindly for several iterations,
  as we demonstrated in Fig.~\ref{fig:unity-filling}.}\BibitemShut {Stop}%
\end{thebibliography}

\end{document}